\documentstyle[aps,preprint]{revtex}
\tighten

\newcommand{\beq}{\begin{equation}}
\newcommand{\eeq}{\end{equation}}
\newcommand{\bea}{\begin{eqnarray}}
\newcommand{\eea}{\end{eqnarray}}

\newcommand{\junk}[1]{}
\def\<{\langle}
\def\>{\rangle}
\def\d{\partial}
\def\+{\dagger}

\def\U1A{U(1)$_{\rm A}$}
\def\LambdaQCD{\Lambda_{\rm QCD}}

\newcommand{\D}{{\cal D}}

\begin{document}

\title{Instanton interactions in dense-matter QCD}
\author{D.T.~Son$^{1,4}$, M.A.~Stephanov$^{2,4}$, and A.R.~Zhitnitsky$^3$}
\address{$^1$ Physics Department, Columbia University, New York, NY 10027, 
USA}
\address{$^2$ Department of Physics, University of Illinois, Chicago, IL 
60607-7059, USA}
\address{$^3$ Department of Physics and Astronomy, University of
British Columbia,\\ Vancouver, BC V6T 1Z1, Canada}
\address{$^4$ RIKEN-BNL Research Center, Brookhaven National Laboratory,
Upton, NY 11973, USA}
\maketitle

\begin{abstract}
A Coulomb gas representation of dense-matter QCD
is derived from a dual transformation of the 
low-energy effective Lagrangian.
The point-like charges $Q=\pm1$ of the gas are identified with 
the instantons and anti-instantons of such topological charges.
An instanton repels another instanton with the same force as
it attracts an anti-instanton, in contrast to the semiclassical
interaction.
\end{abstract}




\section{Introduction and overview}

Instantons have been argued to play an important role in the
nonperturbative dynamics of QCD.  Being classical solutions to the
field equations, instantons provide the starting point for a
phenomenologically successful semiclassical approach to QCD
\cite{SchaeferShuryak}.  Yet the instanton calculus, as this approach
is known, lacks a controllably small parameter which would put it on a
firmer theoretical ground.  The most natural parameter --- the
diluteness of the instanton gas --- is numerically small in QCD
vacuum, but not to the extent that the interaction between the
instantons can be neglected.  The instanton ``gas'' is more similar to
a liquid than a dilute gas.

Given its importance for the QCD vacuum, it is interesting to study
instanton interaction in regimes where the instanton gas can be made
arbitrarily dilute.  In this paper we perform such a study for the
regime of very high matter density, where the diluteness is controlled
by the large ratio of the baryon chemical potential $\mu$ to
$\Lambda_{\rm QCD}$.  We find that in this regime the 
instantons ($I$) and anti-instantons ($\bar I$) interact with each
other via a four-dimensional Coulomb potential generated by their topological
charges.

Our approach is different from the standard semiclassical ones.
A standard instanton calculation starts from the
classical instanton gas and proceeds to calculating quantum corrections. 
In this approach the $II$ and $I\bar I$ 
interactions start as fundamentally different: at the classical level an
instanton interacts with anti-instantons, but not with another instanton.
In contrast, our approach incorporates quantum effects from the very
beginning.  Our starting point
is the low-energy effective Lagrangian for
high-density QCD with two massless quark flavors \cite{ssz}:
\begin{equation}
  L = f^2 [(\d_0\varphi)^2 - u^2 (\d_i\varphi)^2]
   + a \mu^2\Delta^2\cos(\varphi-\theta) \, .
  \label{cos}
\end{equation}
This Lagrangian describes the pseudoscalar isoscalar excitations, $\varphi$
(similar to the $\eta$ meson),
and the dependence of ground state energy on the QCD vacuum angle $\theta$.
Using a well-known formal trick, we rewrite 
the partition function of the theory (\ref{cos}) as that of a Coulomb
gas.  We then notice that $\theta$ plays the role of an external
(electrostatic) potential for this Coulomb gas.  Since, in QCD,
the $\theta$ angle is conjugate to the topological charge, we
identify the constituents of the Coulomb gas with instantons and 
anti-instantons.
In this manner, we obtain information about instanton interaction at
the quantum level, which is not directly available from semiclassical 
calculations.
By comparing the size of the
instanton to the average inter-particle distance in the gas,
we check that the instanton plasma is indeed
dilute for sufficiently large $\mu$. We also
check that the Debye volume contains a large number of instantons, which
means that the instanton plasma is weakly nonideal.


\section{Effective Lagrangian and theta dependence of high-density QCD}
\label{sec:efflag}


Let us briefly recall how Eq.\ (\ref{cos}) is derived (more details
can be found in Ref.\ \cite{ssz}).  The ground state of high-density
QCD with $N_f=2$ light quark flavors ($u$ and $d$) is
a color-superconducting phase with two condensates of left- and right-handed
quarks \cite{colorSC,2SC,2SC2}, 
\begin{eqnarray}
  \<q^{ia}_{L\alpha} q^{jb}_{L\beta} \>^* &=&  
  \epsilon_{\alpha\beta} \epsilon^{ij}\epsilon^{abc} X^c \, ,
  \nonumber \\
  \<q^{ia}_{R\alpha} q^{jb}_{R\beta} \>^* &=&  
  \epsilon_{\alpha\beta} \epsilon^{ij}\epsilon^{abc}Y^c \, .
  \label{2flavor}
\end{eqnarray}
Here the structure of the condensates in spin ($\alpha,\beta$), flavor 
($i,j$) and color ($a,b$) indices is written explicitly.  In the ground
state $X^c$ and $Y^c$ have definite length (see below) and are parallel.
If quarks are massless, then
at the perturbative level there is a degeneracy of the ground state
with respect to
the relative U(1) phase between $X^c$ and $Y^c$, which means that the 
\U1A symmetry of the classical QCD Lagrangian is
spontaneously broken.  The effective Lagrangian of the Goldstone boson
arising from the breaking of \U1A has the following form
\begin{equation}
  L = f^2 [(\d_0\varphi)^2 - u^2 (\d_i\varphi)^2] \, .
  \label{Leff0}
\end{equation}
Since heavier meson excitations have masses of the order of the
superconducting gap $\Delta$,
this Lagrangian is applicable only for momenta much smaller than
$\Delta$.
For large chemical potentials $\mu\gg\LambdaQCD$, the 
leading perturbative values for $f$ and $u$ are \cite{inverse,BBS}:
\begin{equation}
  f^2 = {\mu^2\over 8\pi^2} \, , \qquad  u^2 = {1\over3} \, .
\end{equation}

The field $\varphi$, canonically normalized,
 is the field for the $\eta$ boson:
$\eta=\sqrt2 f\varphi$.
Strictly speaking, $\eta$ is not the lightest mode
in this theory. There is an unbroken color SU(2) sector which
confines at a length scale exponentially larger \cite{RSS}
than the scales we consider in this paper. 
However, the coupling of $\varphi$ to this sector is suppressed
by a power of $\Delta/f$ since the coupling has the form
$(\eta/f) F\widetilde F$. Such a sector is absent for the $N_f=3$
case briefly discussed in Sec.\ \ref{sec:conclusion}.

Instantons add a contribution to (\ref{Leff0}) 
which is given by\cite{ssz}:
\begin{equation}
  V_{\rm inst}(\varphi) = -a \mu^2\Delta^2\cos(\varphi-\theta) \, .
  \label{Vinst}
\end{equation}
This form is uniquely fixed by the anomaly-mediated relation 
of chiral rotations to $\theta$ rotations, and
by the fact that multi-instanton contributions are suppressed
at large $\mu$.  The calculation of the dimensionless function 
$a(\mu)$ is reviewed below.  What is important at this moment is that
$a$ vanishes in the limit $\mu\to\infty$, which means that the mass
of the Goldstone boson becomes much less than $\Delta$ at large $\mu$.
This is the necessary condition for the effective theory (\ref{cos}),
which contains only one field $\varphi$, to have a region of validity. 

If the quark masses, $m_u$ and $m_d$, are non-zero, a new contribution
appears in the effective Lagrangian (\ref{cos}). To lowest order in
the masses its dependence on $\varphi$ is also fixed by symmetry:
\beq
  V_{\rm mass} = - b\, m_u m_d\, \Delta^2 \cos\varphi \, ,
  \label{Vmass}
\eeq
and the coefficient $b\sim 1$ can be calculated using methods
of Ref.\ \cite{inverse}.

The $\theta$ parameter in Eq.\ (\ref{Vinst}) is the fundamental parameter
of the theory which is equal to, or at least very close to, 
zero in our real world.
However, we would like to keep this parameter explicit
because it plays a crucial role in our further arguments.  We note here
that by minimizing
the effective potential $V_{\rm inst}+V_{\rm mass}$ 
with respect to $\varphi$
one can find the theta dependence of the ground state energy.
For small $m_um_d \ll a\mu^2$,
\beq
\label{Evac}
  E_{\rm vac}(\theta)\simeq -a\mu^2\Delta^2-b\,m_um_d\Delta^2\cos\theta,
\eeq
in agreement with the fact that massless fermions suppress the topological
susceptibility (and the theta dependence).

The calculation of $V_{\rm inst}(\varphi)$ can be found in Ref.\ \cite{ssz}.
The starting point is the
instanton-induced effective four-fermion interaction 
\cite{tHooft,SVZ,SchaeferShuryak},
\begin{eqnarray}
   L_{\rm inst} &=& e^{i\theta}\int\!d\rho\, n_0(\rho) 
  \biggl({4\over3}\pi^2\rho^3\biggr)^2 \biggl\{
  (\bar u_R u_L)(\bar d_R d_L) \nonumber\\
  & & + {3\over32} \biggl[ (\bar u_R\lambda^a u_L)(\bar d_R\lambda^a d_L)
  - {3\over4}(\bar u_R\sigma_{\mu\nu}\lambda^a u_L)
    (\bar d_R\sigma_{\mu\nu}\lambda^a d_L) \biggr]
  \biggr\}  + {\rm H.c.} \, .
  \label{inst_vertex}
\end{eqnarray}
By taking the average of Eq.\ (\ref{inst_vertex}) over the
superconducting state (\ref{2flavor}), one finds $V_{\rm inst}$, and
confirms that it is proportional to $\cos(\varphi -\theta)$ 
as in Eq.\ (\ref{Vinst}).  Taking the average of (\ref{inst_vertex}) 
in the superconducting ground state, where
\begin{equation}
  |X| = |Y|  = {3\over 2\sqrt{2}\pi} {\mu^2\Delta\over g} \, ,
\end{equation}
we find
\begin{equation}
  V_{\rm inst}(\varphi) = -\int\!d\rho\, n_0(\rho)
  \biggl({4\over3}\pi^2\rho^3\biggr)^2 12|X|^2\cos(\varphi-\theta) \, ,
\label{Vinst2}
\end{equation}
where $n_0(\rho)$ is the instanton (size) density at finite chemical
potential, which is given by \cite{Shuryak_mu,Carvalho,SchaeferShuryak}
\begin{equation}
\label{n0}
  n_0(\rho) = {0.466 e^{-1.679N_c} 1.34^{N_f}\over(N_c-1)!(N_c-2)!}
  \biggl({8\pi^2\over g^2}\biggr)^{2N_c} \rho^{-5}
  \exp\biggl(-{8\pi^2\over g^2(\rho)}\biggr) e^{-N_f\mu^2\rho^2} \, .
\end{equation}
Taking the $\rho$ integration in Eq.\ (\ref{Vinst2}), we find
\begin{equation}
  a = 5 \times 10^4 \biggl(\ln{\mu\over\LambdaQCD}\biggr)^7
  \biggl({\LambdaQCD\over\mu}\biggr)^{29/3} \, .
  \label{amu}
\end{equation}
Thus $a\to0$ when $\mu\to\infty$, so at sufficiently large $\mu$
the instanton calculations are under analytical control.

\section{Coulomb Gas Representation}
\label{sec:cg}

The effective low energy dense-QCD Lagrangian (\ref{cos})
is the sine-Gordon (SG) Lagrangian.
Many of the special properties of the SG theory apply.
One of these properties is the existence of a kink-type solution, 
corresponding to the domain wall in four dimensions \cite{ssz}.
Another important property is  the admittance of a
Coulomb gas (CG) representation for the partition function.  
Such a representation was used previously \cite{JZ} to argue that, 
in zero-density QCD, the
low energy effective Lagrangian with \U1A anomaly
 represents the  dual form of instanton
contribution to the partition function.  However, QCD at zero density is
in a
nonperturbative regime where no theoretical control is possible \cite{JZ}. 
In the present paper all calculations are under theoretical
control and, therefore, some reliable and precise statements can be made.


The mapping between the SG theory and its CG representation is
well known. All we need to do is reverse the derivation
of SG functional representation of the CG in Ref.\ \cite{Po77}.
The partition function corresponding to
the Lagrangian (\ref{cos}) is given by\footnote{To be precise,
the path integral in Eq.\ (\ref{path_int}) should be understood as an
integral over {\em low-momentum} modes of $\varphi$ only.  The upper
limit of the momentum of $\varphi$ is the ultraviolet cutoff of
the effective Lagrangian (\ref{cos}), which should be taken as some
scale smaller than $\Delta$.  Only tree graphs contribute to $Z$ so
there is no dependence on the precise value of the cutoff.}
\beq
\label{Z}
Z = \int \D \varphi\,  e^{-\int d^3x\,d\tau\,L_E} 
=  \int \D \varphi \ 
e^{ - f^2u \int d^4x (\partial \varphi)^2}\,
e^{ \lambda \int d^4x \cos(\varphi(x)-\theta)} \, ,
  \label{path_int}
\eeq
where we introduced
\beq
\label{lambda}
\lambda \equiv {a\mu^2\Delta^2\over u} \, .
\eeq
$L_E$ is the Euclidean space Lagrangian. The imaginary
time $\tau$ is rescaled to bring the kinetic term into the
Euclidean invariant form in new coordinates with $x_0=u\tau$.
Leaving alone the integration over $\varphi(x)$ for the moment,
we expand the last exponent in Eq.\ (\ref{Z}), represent the
cosine as a sum of two exponents and perform the binomial expansion:
\bea
\label{seriesexp}
{\rm e}^{ \lambda \int d^4x \cos(\varphi(x)-\theta)}
&=&
\sum_{M =0}^\infty \frac{(\lambda/2)^M}{M!} 
\left( \int d^4x\sum_{Q =\pm1}\,e^{iQ(\varphi(x)-\theta)} 
\right)^M 
\nonumber\\
&=&
\sum_{M_\pm=0}^\infty \frac{(\lambda/2)^M}{M_+!M_-!}  
\int d^4x_1 \ldots \int d^4x_M\ 
e^{i\sum_{a=0}^M Q_a(\varphi(x_a)-\theta)}\ .
\eea
The last sum is over all possible sets of $M_+$ positive and $M_-$ 
negative charges $Q_a=\pm1$. The last line in Eq.\ (\ref{seriesexp})
is a classical partition function of an ideal gas of $M=M_++M_-$ identical 
(except for charge) particles of charges $+1$ or $-1$ 
placed in an external potential given by $i(\theta-\varphi(x))$. 
It is easy to see
that (for a constant or slowly varying potential) the average 
number of these particle per unit of 4-volume $\langle M \rangle/V_4$, 
i.e., the density, is equal to $\lambda$. 
Thus making $\lambda$ small one can make the gas
arbitrarily dilute.

While $\theta$ can indeed be viewed as an external potential for the gas
(\ref{seriesexp}), $\varphi(x)$ is a dynamical variable,
since it fluctuates as signified by the path integration in (\ref{Z}).
For each term in (\ref{seriesexp}) the path integral is Gaussian and
can be easily taken:
\beq
\label{intphi}
\int \D \varphi\, e^{ - f^2u \int d^4x (\partial \varphi)^2}\,
e^{i\sum_{a=0}^M Q_a(\varphi(x_a)-\theta)}
=
e^{-i\theta\sum_{a=0}^M Q_a}\,
e^{-{1\over 2f^2u}\sum_{a>b=0}^M  Q_aQ_b
G(x_a-x_b)}\ .
\eeq
We see that, for a given configuration of charges $Q_a$, $-i\varphi(x)$
is the Coulomb potential created by such distribution.\footnote{
One notices that the term $a=b$ in the double sum (\ref{intphi}) is dropped.
This is the self-interaction of each charge. It would renormalize the
fugacity $\lambda$ by a factor $\exp(-G(0)/(f^2u))$. This
factor should be dropped as it represents contribution of very
short wavelength fluctuations of $\varphi$.
Such fluctuations have to be cutoff at the scale $1/\Delta$.
The self-energy of the
charges comes from a much smaller scale, of order $1/\mu$, which is
already calculated and contained in $a$.
}
The function $G(x)$ is the solution of the four-dimensional Poisson
equation with a point source (the inverse of $-\partial^2$):
\beq
G(x_a - x_b) = {1\over 4\pi^2 (x_a-x_b)^2} \, .
\eeq
Thus we obtain the dual CG representation for the partition function
(\ref{Z}):
\beq
\label{CG}
Z = \sum_{M_\pm=0}^\infty \frac{(\lambda/2)^M}{M_+!M_-!}  
\int d^4x_1 \ldots \int d^4x_M\ 
e^{-i\theta\sum_{a=0}^M Q_a}\,
e^{-{1\over 2f^2u}\sum_{a>b=0}^M  Q_aQ_b
G(x_a-x_b)}\ .
\eeq
The two representations of the partition function (\ref{Z}) and
(\ref{CG}) are equivalent.


\section{Physical Interpretation}
\label{sec:meaning}

The charges $Q_a$ were originally introduced in a rather  formal manner
so that the QCD effective low energy Lagrangian can be written in the
dual CG form (\ref{CG}). However, now the
physical interpretation of these charges becomes clear: since
$Q_{\rm net}\equiv \sum_a Q_a$ is the total charge and it appears in
the action
multiplied be the parameter $\theta$ (see Eq.\ (\ref{CG})), one
concludes that $Q_{\rm net}$ {\it is} the total topological
charge  of a given
configuration.  Indeed, in QCD the $\theta$ parameter appears in the
Lagrangian only in the combination with the topological
charge density $-i\theta G_{\mu\nu}
  {\widetilde G}_{\mu\nu}/(32\pi^2)$.  
It is also quite obvious that  each charge 
$Q_a$ in a 
given configuration should be identified with an
integer topological charge well localized at the point $x_a$. This, 
by definition, 
corresponds to a small instanton positioned at $x_a$.
To corroborate this identification we notice that every particle with
charge $Q_a$ brings along a factor of fugacity $\lambda\sim a$ (\ref{lambda})
which contains the classical one-instanton suppression factor
$\exp(-8\pi^2/ g^2(\rho))$ in the density of instantons (\ref{n0}).


The following hierarchy of scales exists in such an instanton ensemble
for sufficiently large $\mu$.
The typical size of the instantons $\bar\rho \sim \mu^{-1}$
is much smaller than the short-distance cutoff of our effective
low-energy theory, $\Delta^{-1}$.
Therefore,
in our low-energy
description they are represented by $\delta(x-x_a)$ functions.
The average distance between the instantons $\bar r=\lambda^{1/4}=
(\sqrt{a}\mu\Delta)^{-1/2}$ is much larger than both the average size 
of the instantons and the cutoff $\Delta^{-1}$.
The largest scale is
the Debye screening length in the Coulomb gas,
$r_D = (\sqrt{a/2}\mu\Delta/fu)^{-1}\sim (\sqrt{a}\Delta)^{-1}$.
This coincides with the static correlation length of the $\varphi$ field, 
which differs from the rest mass of the Goldstone by a factor of $u$.
In short:
\beq
\begin{array}{ccccccc}
\mbox{(size)}&\ll&\mbox{(cutoff)}&\ll& \mbox{(distance)}&\ll& \mbox{(Debye)}\\
\mu^{-1} &\ll&\Delta^{-1}&\ll&(\sqrt{a}\mu\Delta)^{-1/2}&\ll&(\sqrt{a}\Delta)^{-1}
\end{array}
\eeq
Due to this hierarchy, ensured by large $\mu/\Lambda_{\rm QCD}$, we acquire 
analytical control.

It is also quite interesting that,
although the starting low-energy effective Lagrangian contains only 
a colorless field $\varphi$, we
have ended up with a representation of the partition function in which 
objects carrying color (the instantons, their interactions and
distributions) can be studied.
In particular, from the discussions given above, one can immediately
deduce that $II$ and $I\bar I$
interactions are exactly the same up to a sign 
and are Coulomb-like at large distances.
  
This looks highly nontrivial
since it has long been known  that 
at the semiclassical level
an instanton interacts only
with anti-instantons 
but not with 
another instanton carrying a topological charge of the same sign.
As we demonstrated above it is not true any more at the quantum level.
Indeed, what we have found is that the interactions between 
dressed (as opposed to than bare) instantons and anti-instantons
after one takes into account
their  classical and quantum interactions,
after integration over their all possible sizes and color orientations, after
accounting for the interaction with the background condensates $X$ and
$Y$ in Eq.\ (\ref{2flavor}) 
(which by itself is   nonzero  due to the same (anti)instantons 
as well as due to purely perturbative interactions),
must become very simple at large distances as 
explicitly described by Eq.\ (\ref{CG}).
One could be really amazed how the problem which
looks  so complicated in terms of the original bare (anti)instantons, 
becomes so simple in terms of the dressed (anti)instantons
when all integrations over all possible sizes, color orientations
and interactions with background fields are properly accounted for!

Such a simplification of the interactions is
of course due to the presence of an almost massless
pseudo-Goldstone boson $\eta$ which couples to the topological charge.
When the instanton gas becomes very dilute all semiclassical 
interactions (due to zero
modes) cannot contribute much, since they fall off with distance
faster then the Coulomb interaction mediated by $\eta$. 
On the other
hand, when the instanton density increases (at lower baryon
densities), the Coulomb interaction becomes more screened and, as
the Debye length becomes comparable to the inter-instanton distances,
we lose analytical control.

Another interesting observation is that the Gauss law manifests itself
in the CG partition function by suppressing all configurations with
nonzero net charge $Q_{\rm net}$. (The weight of such configurations
is suppressed by $\exp(-\# Q_{\rm net}^2/L^2)$, 
where $L$ is the linear size of the system. This restricts
fluctuations of $Q_{\rm net}$ to $O(L)$, negligible compared to
normal thermodynamic size $O(L^2)$.) In QCD vacuum 
this corresponds to the suppression
of the topological charge fluctuations in the chiral limit.
Nonzero quark masses give the $\eta$ particle a mass, turning
the long-range Coulomb
interaction into a Yukawa one, thus unfreezing fluctuations of $Q_{\rm net}$,
in agreement with a well known result that topological charge
susceptibility is proportional to quark masses (to the product
$m_um_d$ in our case (\ref{Evac})).

\section{Conclusions and outlook}
\label{sec:conclusion}

In this paper we study a regime of QCD where the instanton
dynamics acquires analytical control --- the high baryon density regime.
We showed that the partition function
(\ref{CG}) of the Coulomb gas of instantons is the dual
representation of low-energy high-density QCD (\ref{cos}) 
as far as the ground state structure, in particular,
its theta dependence, is concerned. 
We discuss the correspondence between these
two representations. The most
important physical result
is a rather straightforward identification
of the charges $Q_a$ from the statistical ensemble
(\ref{CG}) with the well-known
BPST-instantons\cite{BPST,Atiyah} which have been under
intensive study since 1970's.
We have explicitly demonstrated that at large distances
instantons and
anti-instantons interact like point-like charged particles.
This is in contrast with standard semiclassical approximation
when $II$ and $I\bar I$ interactions
are very different from each other.

A similar correspondence between a formal expression for a 
statistical ensemble of particles and the
instanton quarks suspected long ago\cite{Belavin}
was conjectured recently\cite{JZ} based
on the analysis of the multi-instanton measure.  As we have said, 
no theoretical
control was possible in Ref.\ \cite{JZ} to justify that conjecture.
Only in the high-density regime considered 
in the present paper the  relation between
$\eta$ physics  and dual representation of the ensemble
of instantons  can be made precise.  The properties of the instanton
ensemble
which we find should be of interest for numerical instanton
simulations.

It is possible to generalize our results to the color-flavor-locking 
(CFL) state of $N_f=3$ QCD \cite{CFL}.  The \U1A
symmetry is also spontaneously broken in this case.  The role of the
$\eta$ boson is played by the $\eta'$ meson, which is also light
at high densities \cite{CFL,inverse}. The instanton-induced $\eta'$ 
potential has a form similar to (\ref{Vinst}) \cite{MT,ssz}:
\begin{equation}
  V_{\rm inst}(\varphi) = -a'\cdot
  \mu^2\Delta^2\cos(\varphi-\theta) \, ,
\end{equation}
where the evaluation of dimensionless function $a'$ is very much the same
as our calculation of $a$ in Sec.\ \ref{sec:efflag} or Ref.\ \cite{ssz}.
We need to insert an extra factor
$m_s\rho$ into (\ref{Vinst2}) and use $N_f=3$ in $g(\rho)$
in (\ref{n0}) to find:
\beq
a'  = 7 \times 10^3 \left(m_s\over\mu\right)
	\biggl(\ln{\mu\over\LambdaQCD}\biggr)^7
  \biggl({\LambdaQCD\over\mu}\biggr)^{9} \, .
  \label{a'mu}
\eeq

Finally, the same analysis can be also carried in the case of QCD
at large isospin density $\mu_I$ \cite{isospin}. The instanton interactions
can be also shown to be Coulomb-like and the diluteness
of the gas is achieved at large $\mu_I$.
In contrast to finite baryon density case, such a system
can be simulated on the lattice today.



\acknowledgements

We are indebt to D.~Kharzeev, L.~McLerran, R.~Pisarski,
T.~Sch\"afer, and E.~Shuryak
for discussions.  We are thankful to the INT at University of
Washington for the organization of the workshop ``QCD at finite baryon
density'' which made this collaboration possible.  DTS and MAS thank
RIKEN, Brookhaven National Laboratory, and U.S.\ Department of Energy
[DE-AC02-98CH10886] for providing the facilities essential for the
completion of this work.  The work of DTS is supported, in part, by a
DOE OJI Award.  AZ is supported in part by the National Science and
Engineering Research Council of Canada.

   \end{document}